\documentclass[aps,physrev,10pt,twocolumn,superscriptaddress,reprint]{revtex4-1}
\usepackage{hyperref}
\usepackage{amsmath}
\usepackage{graphicx}
\usepackage{epstopdf}
\usepackage{mathrsfs}
\usepackage{placeins}

\newcommand{\pdiff}[2]{\frac{\partial #1}{\partial #2}}

\graphicspath{{./figures/}}

\begin{document}

\title{Self-driven oscillation in Coulomb blockaded suspended carbon nanotubes}
\date{\today}
\author{Kyle Willick}
\affiliation{Institute for Quantum Computing, University of Waterloo, Waterloo, Ontario, Canada N2L 3G1}
\affiliation{Waterloo Institute for Nanotechnology, University of Waterloo, Waterloo, Ontario, Canada N2L 3G1}
\affiliation{Department of Physics and Astronomy, University of Waterloo, Waterloo, Ontario, Canada N2L 3G1}
\author{Jonathan Baugh}
\email[Contact: ]{baugh@uwaterloo.ca}
\affiliation{Institute for Quantum Computing, University of Waterloo, Waterloo, Ontario, Canada N2L 3G1}
\affiliation{Waterloo Institute for Nanotechnology, University of Waterloo, Waterloo, Ontario, Canada N2L 3G1}
\affiliation{Department of Chemistry, University of Waterloo, Waterloo, Ontario, Canada N2L 3G1}

\begin{abstract}
Suspended carbon nanotubes are known to support self-driven oscillations due to electromechanical feedback under certain conditions, including low temperatures and high mechanical quality factors. Prior reports identified signatures of such oscillations in Kondo or high-bias transport regimes. Here, we observe self-driven oscillations that give rise to significant conduction in normally Coulomb-blockaded low-bias transport. Using a master equation model, the self-driving is shown to result from strongly energy-dependent electron tunneling, and the dependencies of transport features on bias, gate voltage, and temperature are well reproduced.
\end{abstract}

\maketitle

\begin{figure}
\includegraphics[width=\linewidth]{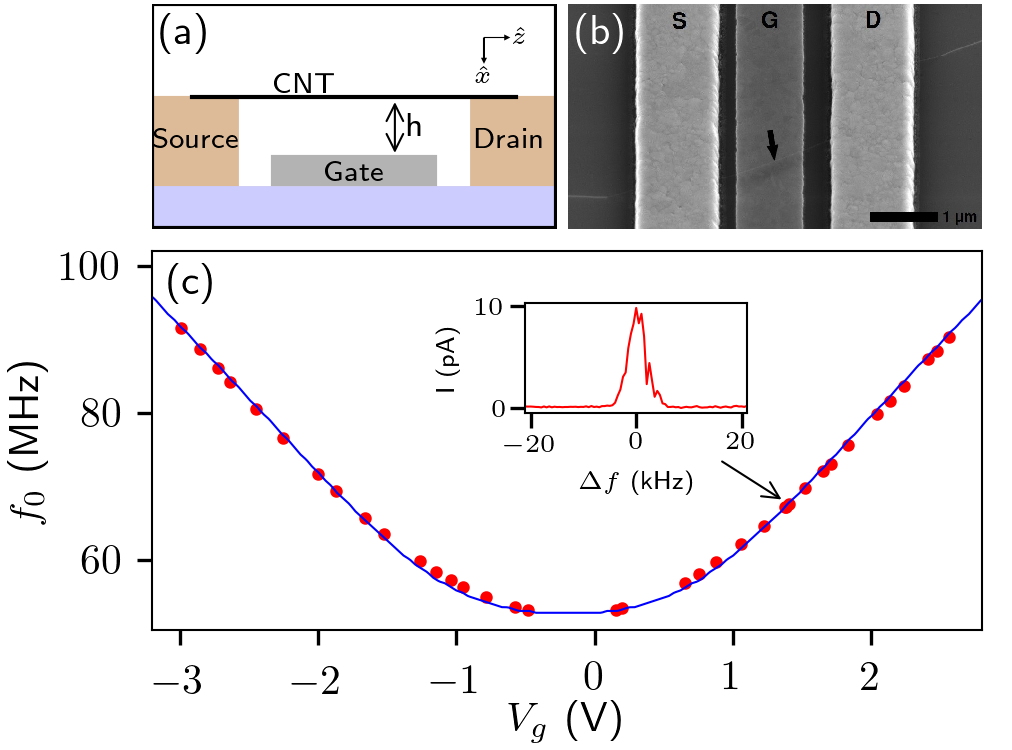}
\caption{\textbf{a}. Schematic of the suspended CNT transistor device. The CNT is contacted by Ti/Au source and drain contacts, and suspended over a Ni gate electrode. The CNT/gate separation is approximately 340~nm, with a nominal suspended length of $2\:\mu\text{m}$. \textbf{b}. SEM image of a device similar to the one measured. The arrow highlights the CNT location. \textbf{c}. Mechanical resonance frequency as a function of applied DC gate voltage, measured at $30\:\text{mK}$. The red markers indicate experimental values, and the blue line is a theoretical fit using an Euler-Bernoulli beam model. Fit parameters include a CNT diameter of $1.9\:\text{nm}$, suspended length of $2.2\:\mu\text{m}$, and residual compression of $T_0 =  35\:\text{pN}$. The inset shows a resonance peak measured by the Coulomb rectification technique. The peak width gives a lower bound on the mechanical quality factor, $Q > 10^4$. }
\label{fig:dev}
\end{figure}

\begin{figure}
\includegraphics[width=\linewidth]{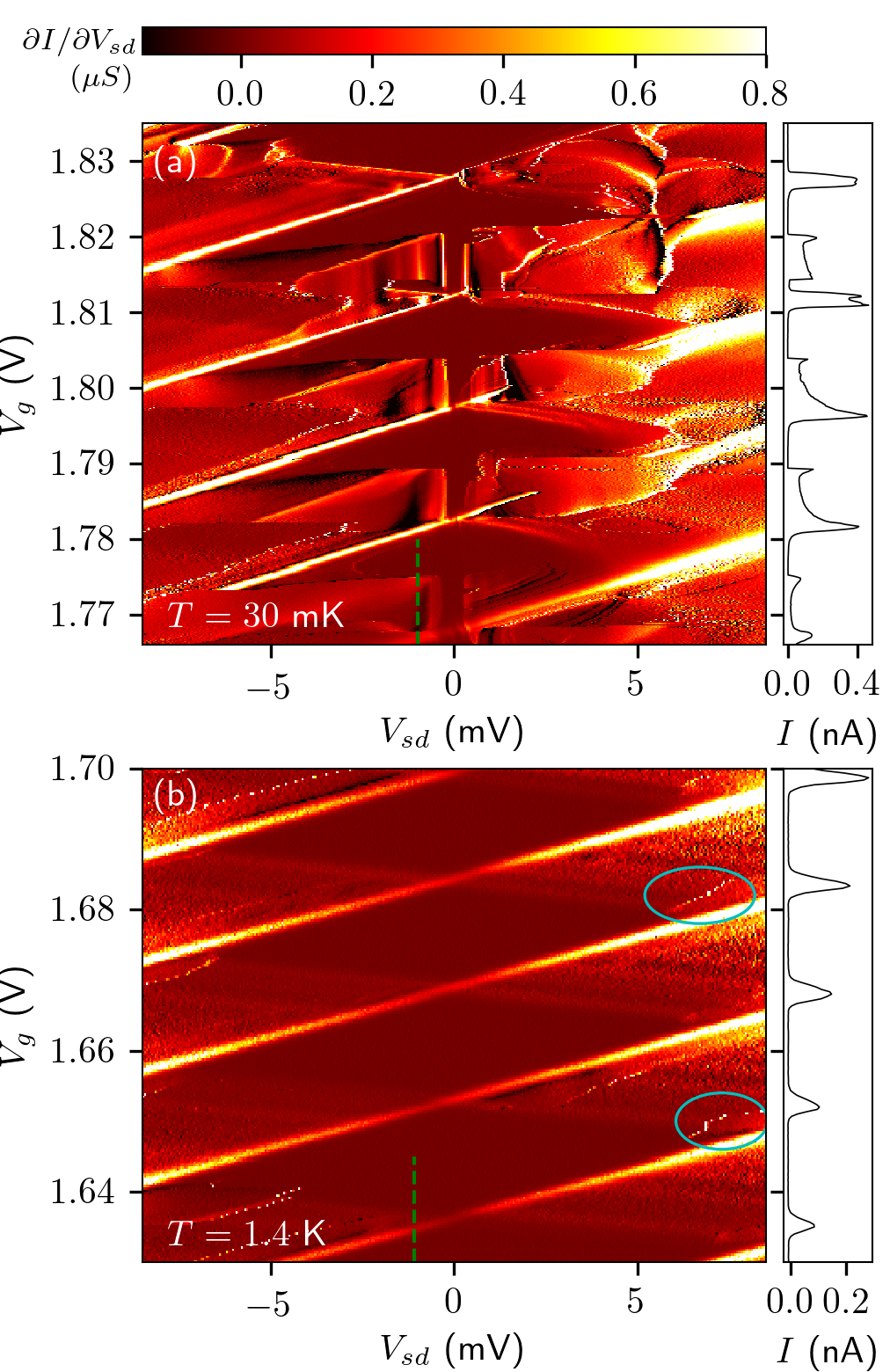}
\caption{\textbf{a}. Differential conductance, $\pdiff{I}{V_{sd}}$, as a function of DC gate voltage and bias for a suspended CNT device at $T = 30\:\text{mK}$. Large intrusions of finite conductance are observed within several nominally zero-conductance Coulomb diamonds. A linescan shown at right gives the current, $I$, measured at bias $V_{sd} = -1\:\text{mV}$, in which these conductance features appear as extended side lobes on the normal Coulomb peaks. \textbf{b}. Differential conductance and current linescan of the same device measured at $T = 1.4\:\text{K}$, in which the normal Coulomb diamonds and peaks are observed. Sharp conductance ridges are seen at high bias outside of the diamonds (two examples highlighted by blue circles). This indicates that self-driven oscillations are present at 1.4~K, but do not produce current inside the diamonds at that temperature.}
\label{fig2}
\end{figure}

Nanoscale electro-mechanical resonators based on suspended carbon nanotubes (CNTs) have achieved state-of-the-art sensitivity in mass and force detection \cite{Chaste2012, Moser2013}, owing to the small mass and high quality (Q) factors of CNTs. Under certain conditions, a positive feedback between the tunneling of charge carriers on/off the CNT and mechanical motion can lead to self-driven oscillations \cite{Usmani2007}. The study of self-driving phenomena is itself fundamentally interesting, with implications for mass/force sensing, and the underlying carrier-motion coupling has potential application for mechanical cooling \cite{Urgell2019}.
Self-driven oscillations were first observed in the electronic transport of suspended CNT transistors with high mechanical Q factors \cite{Steele2009}. Further studies verified the mechanical nature of these transport signatures \cite{Schmid2012,Schmid2015}. Similar features were recently observed in Kondo \cite{Urgell2019} and high-bias tunnelling transport \cite{wen2020coherent}, where advanced readout techniques confirmed the large amplitude and bistability of self-driven states. In these prior studies, the transport signatures of self-driven oscillations appeared as instabilities in the current-carrying transport regime. In this letter, we report the experimental observation at sub-Kelvin temperatures of self-driven oscillations that produce significant nonzero current in the otherwise Coulomb-blockaded, zero-current regime. The feedback system is modeled using a master equation approach that includes energy-dependent tunneling, and excellent qualitative agreement between theory and experiment is obtained. It is concluded that energy-dependent tunneling leading to negative damping of the mechanical motion is a necessary condition for producing these oscillation states.  \\
\indent The suspended nanotube devices were fabricated by a process in which the CNT is grown separately from the device wafer, then transferred onto predefined contacts by stamping \cite{Wen2018, waissman2013realization}. The device consists of 400~nm thick Ti/Au source/drain electrodes on a SiO$_2$/Si substrate. A single 60~nm thick Ni gate electrode is placed within the gap between the contacts. Figure \ref{fig:dev}a is a schematic of the device design, and figure \ref{fig:dev}b shows a scanning electron microscope (SEM) image of a device similar to those used in experiments. Experiments were performed at 1.4~K in a pumped He$_4$ cryostat and at 30~mK and 800~mK in an Oxford Instruments DR200 dilution refrigerator. Two devices were measured on the same chip during the same cool-downs, and were found to have similar characteristics and to exhibit qualitatively similar self-driven oscillations. We focus on device 1 in this paper, and provide data from device 2 in Appendix C. \\
\indent The mechanical resonance frequency was measured as a function of gate voltage using a Coulomb rectification technique \cite{Huttel2009_Nano}. The experimental resonance frequencies are shown by the points in figure \ref{fig:dev}c, fit by an Euler-Bernoulli beam model (solid line) of the gate voltage dependence, which gives an estimate of the CNT diameter ($1.9\:\text{nm}$), suspended length ($2.2\:\mu\text{m}$), and residual compression ($35\:\text{pN}$) \cite{Poot2012}. The linewidths of the resonance peaks at fixed gate voltage provide a lower bound for the mechanical quality factor of $Q > 10^4$. Axial field magneto-spectroscopy supports the CNT diameter estimate of approximately $2\:\text{nm}$ (see appendix A).  \\
\indent The self-driven oscillation features are apparent in measurements of the CNT conductance at 30~mK. Figure \ref{fig2}a shows differential conductance versus bias voltage and gate voltage, for a region of electron transport that includes several Coulomb blockade diamonds. Overlaid on the Coulomb diamond structure are finite conduction features that occur within the normally fixed-occupation, zero conductance region. These features do not depend on the measurement sweep direction, and are stable in time. Similar features were observed at other gate voltage ranges (see appendix B).\\
\indent For comparison, figure \ref{fig2}b shows a similar measurement performed on the same device at 1.4~K, for which these additional finite conduction features inside the diamonds are absent. These features were absent in all Coulomb diamonds at 1.4~K, but were seen in a significant fraction of the electron transport diamonds at sub-Kelvin temperatures. These features were not observed in the hole transport for either of the two devices investigated. In figure \ref{fig2}b at 1.4~K, several sharp, high conductance `ridges' can be seen in the high bias tunneling current regime, similar to self-driven oscillation features reported in the literature \cite{wen2020coherent}. The device capacitances and approximate resonant tunnel rates were determined by analysis of the diamond edges. Values obtained for device 1 were $C_g = 12.3\:\text{aF}$, $C_s = 5.6\:\text{aF}$, $C_d = 8.4\:\text{aF}$, $\Gamma_s = 3\:\text{GHz}$, and $\Gamma_d = 8\:\text{GHz}$ (subscripts $g, s, d$ refer to gate, source and drain, respectively). \\
\indent The finite current features within the diamonds observed in figure \ref{fig2}a share the sharp, rough edges of the high-bias self-driven features of \ref{fig2}b. The stretching of these features into the Coulomb diamond regions in figure \ref{fig2}a is reminiscent of the widening of the Coulomb peak width under strong resonant mechanical driving, i.e. Coulomb rectification \cite{Huttel2009_Nano}. The self-driven features are typically asymmetric with respect to the gate voltage and 
indicate a large amplitude of motion. For example, using the device parameters determined above to estimate $\partial{C_g}/\partial{x}$, the finite current observed near $V_g = 1.82\:\text{V}, V_{sd} = -1\:\text{mV}$ corresponds to an approximate oscillation amplitude of $5\:\text{nm}$.\\
\indent To simulate the dot-mechanical system, we consider an extension of the model which first predicted such self-driven oscillations \cite{Usmani2007}. The extension presented here 
considers two electron occupation transitions ($n \rightarrow n+1$ and $n \rightarrow n-1$), and specifically focuses on the region within the Coulomb diamond. The model system is a quantum dot capacitively coupled to a fixed gate and tunnel-coupled to two leads. The dot position relative to the gate is free to vary, with $x$ being the displacement of the position from an initial separation $h$.\\
\indent For $x \ll h$, the chemical potential of the dot as a function of position is given by \cite{Brink2006} $
\mu_{dot}(x) = \frac{e}{C_{tot}} \left(C_g(0) V_g + C_s V_s + C_d V_d\right) + \frac{e}{C_{tot}}\pdiff{C_g}{x} V_g x$, where 
$C_{tot} = C_g + C_s + C_d$. Also, the electric force between the gate and the dot is given by $F_{dot} \approx \frac{1}{2}\pdiff{C_g}{x}V_g^2 - \pdiff{C_g}{x} \frac{V_g}{C_{tot}} \left( C_g V_g + C_s V_s + C_d V_d \right) + \frac{e}{C_{tot}} \pdiff{C_g}{x} V_g \langle N \rangle$, where $\langle N \rangle$ is the average electron occupation of the dot. Therefore, the electromechanical coupling is parameterized by the coupling energy $F = \frac{e}{C_{tot}}\pdiff{C_g}{x}V_g$.\\
\indent For the CNT system considered here, several simplifying assumptions based on the relevant energy scales are made. First, the mechanical motion of the dot is taken to be in the classical regime ($\hbar\omega_0 \ll k_bT$), so it is well described by a classical position and velocity~($x$, $v$). Second, the tunnel rates into the dot are much faster than the time scales of mechanical motion ($\Gamma \gg \omega_0$), so that individual tunneling events occur at approximately fixed CNT positions, with only small displacements between successive tunneling events. Finally, the mechanical resonance is only weakly damped ($Q \gg 1$) so that the mechanical state changes little between successive oscillation periods. \\
\indent The state of the system can be described by joint probability distributions, $P_n(x,v,t)$, where $P_n$ is the probability distribution of variables $(x,v,t)$ for electron occupation $n$. Consider an initial state within the Coulomb diamond so that the average electron occupation is an integer, $\langle N \rangle = n$. Then the probability distributions obeys the coupled equations \cite{Armour2004, Usmani2007}
\begin{widetext}
\begin{equation}
\begin{split}
\mathscr{L}{P_{n-1}} - \frac{F}{M} \pdiff{P_{n-1}}{v} = -\Gamma_{n-1 \rightarrow n} P_{n-1} + \Gamma_{n \rightarrow n-1} P_n \\
\mathscr{L}{P_{n}} = \Gamma_{n-1 \rightarrow n} P_{n-1} - \left(\Gamma_{n\rightarrow n-1} + \Gamma_{n \rightarrow n+1}\right)P_n + \Gamma_{n+1 \rightarrow n} P_{n+1} \\
\mathscr{L}{P_{n+1}}+ \frac{F}{M} \pdiff{P_{n+1}}{v} = \Gamma_{n \rightarrow n+1} P_n - \Gamma_{n+1 \rightarrow n} P_{n+1}
\end{split}
\end{equation}
\end{widetext}
where $\Gamma_{a \rightarrow b} = \Gamma^s_{a \rightarrow b} + \Gamma^d_{a \rightarrow b}$ is the tunnel rate from state $a$ to $b$, and $\mathscr{L}$ describes the mean-coordinate evolution of an oscillator with frequency $\omega_0$ and quality factor $Q$ \cite{Armour2004}
\begin{equation}
\mathscr{L}P_n = \pdiff{P_n}{t} + v\pdiff{P_n}{x} - \omega_0^2 x \pdiff{P_n}{v} - \frac{\omega_0}{Q} \pdiff{vP_n}{v}
\end{equation}

Under the assumptions $\Gamma \gg \omega_0$ and $e V_{s,d} < e^2 / C_{tot}$, the tunneling events for $n \leftrightarrow n+1$ and $n \leftrightarrow n-1$ are well separated in time, so that one of $P_{n-1}$ or $P_{n+1}$ will be zero when evaluating the above expressions for any $(x, v, t)$. We make the further assumption that the instantaneous occupation of the quantum dot is a small perturbation from the occupation which would occur with $v = 0$. That is, the probabilities can be written $P_{n} = N_{a \leftrightarrow b} P + \delta P$ where $\delta P$ is small and $N_{a \leftrightarrow b} = \frac{\Gamma_{a \rightarrow b}}{\Gamma_{a \leftrightarrow b}}$ with $\Gamma_{a \leftrightarrow b} = \Gamma_{a \rightarrow b} + \Gamma_{b \rightarrow a}$.\\
\indent Following a similar analysis to that presented in the literature \cite{Blanter2004, Usmani2007}, and additionally noting the phase-averaged contribution of terms of the form $f\left(A \sin(\phi)\right) \cos(\phi)$ will be zero, the probability distribution for the oscillator to have mechanical energy E satisfies
\begin{small}
\begin{equation} \label{eq:dPdT}
\pdiff{P}{t} = \pdiff{}{E}\left(2 E \langle \gamma(x) \cos^2(\phi) \rangle P + 2 E M \langle D(x) \cos^2(\phi)  \rangle \pdiff{P}{E}\right)
\end{equation}
\end{small}
where $\phi$ is the mechanical phase, 
$\langle * \rangle = \frac{1}{2\pi} \int_0^{2\pi} *\, d\phi$ is the average over one oscillation period, and the damping ($\gamma$) and diffusion ($D$) terms are
\begin{small}
\begin{equation*}
\gamma(x) = \frac{F^2}{M} \left(  \frac{1}{\Gamma_{n-1 \leftrightarrow n}} \pdiff{N_{n-1 \leftrightarrow n}}{\mu} + \frac{1}{\Gamma_{n \leftrightarrow n+1}} \pdiff{N_{n \leftrightarrow n+1}}{\mu} \right) + \frac{\omega_0}{Q}
\end{equation*}
\begin{equation*}
D(x) = \frac{F^2}{M^2} \left( \frac{\Gamma_{n-1 \rightarrow n}\Gamma_{n \rightarrow n-1}}{(\Gamma_{n-1 \leftrightarrow n})^2} + \frac{\Gamma_{n \rightarrow n+1}\Gamma_{n+1 \rightarrow n}}{(\Gamma_{n \leftrightarrow n+1})^2}  \right)
\end{equation*}
\end{small}
The steady state solutions to equation (\ref{eq:dPdT}) are
\begin{equation} \label{eq:PE}
P(E) \propto \exp\left(-\int_0^E \frac{\langle \gamma(\sqrt{\frac{2 E'}{k}}\sin(\phi)) \cos^2(\phi) \rangle}{M \langle D(\sqrt{\frac{2 E'}{k}}\sin(\phi)) \cos^2(\phi) \rangle} dE'\right)
\end{equation}

\begin{figure}
\includegraphics[scale=1]{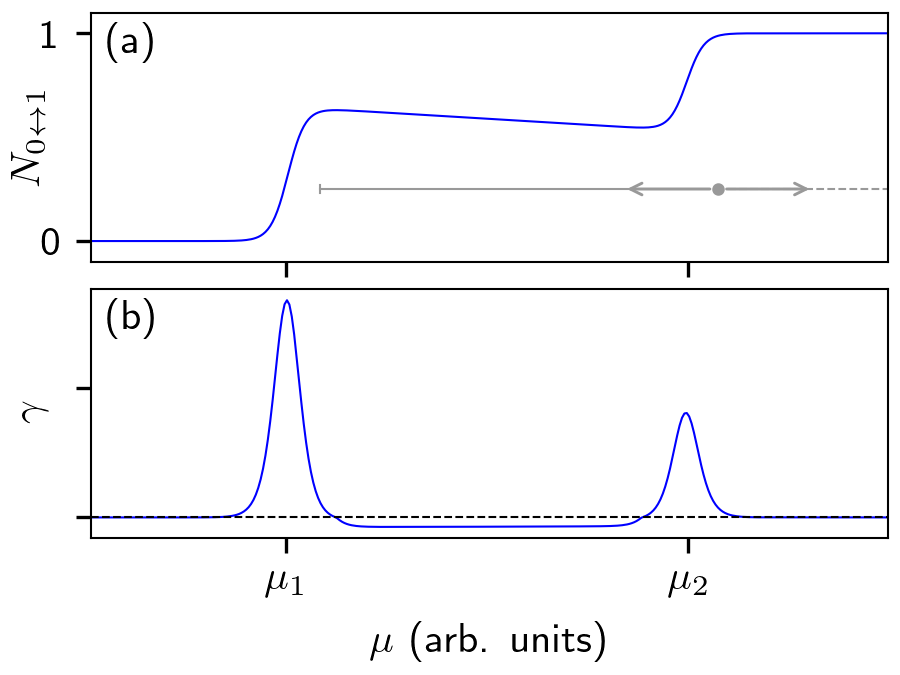}
\caption{Energy dependent tunneling across a charge state transition at finite bias, in which sequential tunneling occurs when the dot chemical potential is within the bias window ($\mu_1 < \mu < \mu_2$). \textbf{a.} For energy dependent tunneling, the average occupation ($N_{0 \leftrightarrow 1}$) is not fixed within this transition. \textbf{b.} This can result in negative damping of the mechanical motion, $\gamma$. In the negative damping regime, oscillations will increase in amplitude until reaching a stable oscillation state. The grey illustration in \textbf{a} depicts a mechanical-dot system near the $\mu_2$ edge entering into this negative damping region and increasing in amplitude until $\gamma$ switches sign near $\mu_1$.}
\label{fig:energyDep}
\end{figure}

For the Coulomb blockade region of interest, equation (\ref{eq:PE}) will always permit a nonzero peak at $E=0$, which corresponds to zero motion. Thus, the regime of interest for self-driven oscillation is when a second probability peak exists at nonzero $E$. In this case, the oscillator / dot system can sustain self-driving at a finite amplitude. The experimental observation of stable current within diamonds in figure \ref{fig2}a indicates occupation of these higher energy states. In the following simulations we consider the case in which the higher energy state is occupied whenever it has a nonzero probability. This could be driven by thermal fluctuations exciting the resonator to a regime of negative damping, which is then rapidly driven to a large amplitude stable oscillation by the mechanism discussed below. \\
\indent For a peak in $P(E)$ to exist at nonzero energy, the integrand in equation \ref{eq:PE} must take negative values for some range. As $D(x) > 0$ for all $x$, a necessary condition for self driven oscillations is then that $\gamma(x) < 0$ for some $x$, which implies
\begin{equation}
\pdiff{N_{a \leftrightarrow b}}{\mu} < - \frac{M \Gamma_{a \leftrightarrow b}}{F^2}\frac{\omega_0}{Q}.
\end{equation}
This requires a high $Q$ mechanical resonator and some energy dependence of the tunnel barrier. An example of the average electron occupation ($N_{a\leftrightarrow b}$) across a finite bias Coulomb peak with energy dependent tunneling is shown in figure~\ref{fig:energyDep}, along with the resultant damping term $\gamma$. Self-driven oscillations are possible when the motion-averaged damping becomes negative. For initial configurations within Coulomb blockade, this occurs when there is a region of negative damping within an adjacent charge state transition, and only small positive damping contributions at the transition edges.\\
\indent Several mechanisms could be responsible for tunnel energy dependence, such as low-wide tunnel barriers \cite{Korotkov1991}, finite density of states in the leads \cite{wang2005carbon}, and non-uniform profile of the CNT potential above the contacts \cite{perello2010anomalous, knoch2008tunneling, gehring2017distinguishing}. Here we take a phenomenological model of the tunnel rates similar to that previously applied to suspended CNTs \cite{Meerwaldt2012}
\begin{equation} \label{eq:tunnel}
\Gamma^{s,d} = \Gamma_0^{s,d} e^{ b^\pm_{s,d} \Delta\mu_{s,d}} f_F(\Delta\mu_{s,d})
\end{equation}
where $\Delta\mu_{s,d} = \left(\mu_{dot} - \mu_{s,d}\right)$, $\mu_{s,d}$ is the energy level of the source/drain contact, $\Gamma_0^{s,d}$ is the resonant tunnel rate at $\Delta\mu_{s,d} = 0$, and $b^\pm_{s,d}$ are fitting parameters describing the energy dependence of the rate from the source/drain contact at negative/positive bias, respectively.
As a final step in the simulation, the current is evaluated using 
\begin{equation*}
I = \langle I(x) \rangle = e  \left\langle \frac{\Gamma^s_{a \rightarrow b}(x) \Gamma^d_{b \rightarrow a}(x) - \Gamma^s_{b \rightarrow a}(x) \Gamma^d_{a \rightarrow b}(x)}{\Gamma_{a \leftrightarrow b}(x)} \right\rangle
\end{equation*} 
where $\{a,b\} = \{n-1,n\},\{n,n+1\}$.\\
\begin{figure}
\centering
\includegraphics[width=0.98\linewidth]{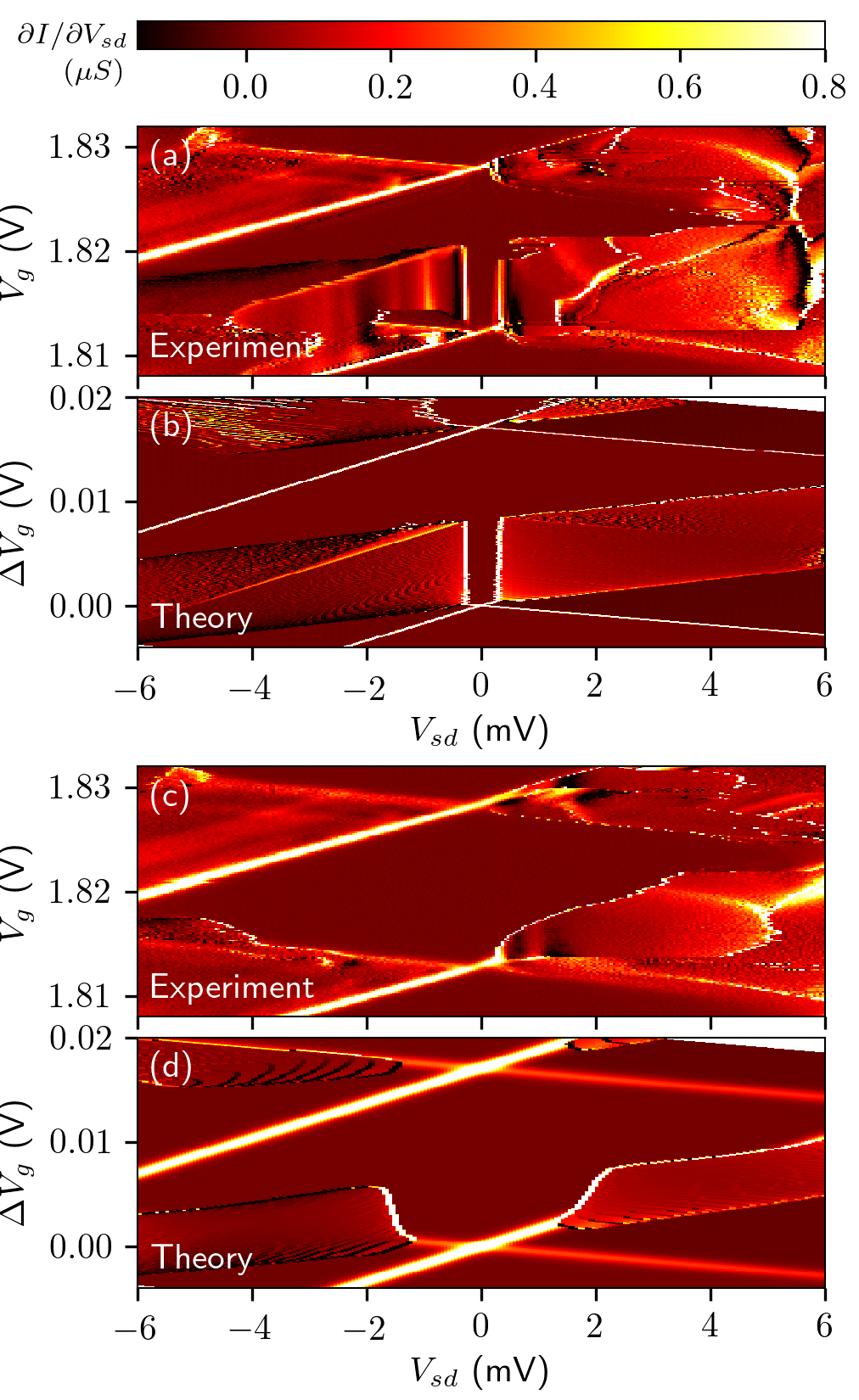}
\caption{\textbf{a}. Experimental differential conductance measured at $T = 30\:\text{mK}$, showing finite conductance due to self-driven oscillations in the Coulomb diamond region between two charge state transitions. \textbf{b}. Simulated differential conductance using the master equation model described in text, with best-fit energy dependent tunneling parameters $b_L^+ = b_R^- = 6\:\text{meV}^{-1}$, $b_R^+ = b_L^- = 0.2\:\text{meV}^{-1}$. \textbf{c}. Differential conductance over the same range as \textbf{a} measured at $T = 800\:\text{mK}$. \textbf{d}. Simulated conductance using the same parameters as \textbf{b} but with $T = 800\:\text{mK}$.}
\label{fig:selfModelFit}
\end{figure}
\indent Figure \ref{fig:selfModelFit}a focuses on one dot occupation level of the differential conductance experiment previously shown in figure \ref{fig2}a. Figure \ref{fig:selfModelFit}b shows the simulated differential conductance using the model, over an equivalent gate voltage range, with the device parameters determined previously and $b_{s,d}^\pm$ parameters determined by fitting.  For the parameters shown, the simulation qualitatively reproduces many of the features observed within the Coulomb diamond in the experiments. Current is supported within the Coulomb diamond by self-driven oscillations of sufficient amplitude to modulate $\mu_{dot}$ past the diamond edge. The self-driven features are absent at low bias, and abruptly terminate at the nearly horizontal mid-line of the Coulomb diamond. At this mid-line, the mechanical oscillations begin to interact with both the $n \rightarrow n-1$ and $n \rightarrow n+1$ transitions simultaneously, leading to large positive damping. The gap at low bias is a result of the thermal broadening of the shoulders (at $\mu_1$ and $\mu_2$ in figure \ref{fig:energyDep}) overtaking the negative slope region of $\langle N \rangle$ as bias is decreased. Also note the appearance of the bright conductance line at negative bias that runs parallel to the opposite Coulomb diamond edge. Here, the self-oscillation due to energy dependence in $n \rightarrow n-1$ tunneling has sufficient amplitude to allow transport through the $n+1$ energy level for a portion of the oscillation period, resulting in a sharp increase in conductance. Some features not captured by the simulations, such as the multiple switching features at positive $V_{sd}$ in figure \ref{fig:selfModelFit}a, may be attributable to higher order mechanical modes or additional electronic transitions that are not considered in the present model.\\
\indent Figures \ref{fig:selfModelFit}c and d show experimental and simulated differential conductance for the same parameter space, at elevated temperature $T = 800\:\text{mK}$. As temperature increases, the self-oscillation conductance feature decreases in extent and requires larger bias voltages. The temperature dependence of the simulated self-oscillation is qualitatively similar, but weaker than in experiment, likely because the simulation does not take into account a temperature dependence of the tunneling parameters, $\Gamma_0$ and $b_{s,d}^\pm$. The presence of self-driven oscillation features within Coulomb diamonds indicates a strong energy dependence of tunneling at the adjacent diamond edge. Reducing the energy dependence parameters, $b_{s,d}^\pm$, decreases the negative damping strength and increases the threshold of bias needed to observe these features. When this threshold exceeds the charging energy, finite current features are no longer observed within the diamond. For the simulation shown in figure \ref{fig:selfModelFit}b, decreasing $b_{s,d}^\pm$ by a factor of 10 completely suppresses the self-oscillation features.\\
\indent The stable, large amplitude oscillations discussed above may interfere with the usual operation of these devices as nano-mechanical force or mass sensors at sub-Kelvin temperatures. However, these oscillations may exhibit narrower linewidths than the intrinsic resonator linewidth \cite{Bennett2006}, and a properly tailored electromechanical coupling can provide a pathway to active cooling of the mechanical state \cite{Urgell2019}. Additionally, with sensitive frequency resolved readout \cite{Wen2018}, self-driven oscillations may be useful for sensitive mass/force detection without the need for high frequency external excitation at the device \cite{Feng2008}. These self-driving devices could also see use for cryogenic compatible RF components. In particular, as conductance can be suppressed by Coulomb blockade for large portions of the mechanical oscillation, these devices can act as voltage tunable RF timing bases with very low intrinsic power dissipation ($< 1\:\text{pW}$).

% Acknowledgements
\vspace{5mm}
\section*{Acknowledgements}
\vspace{-3mm}
We thank Yutian Wen and Bohdan Khromets for useful discussion. This research was undertaken thanks in part to funding from the Canada First Research Excellence Fund, Natural Sciences and Engineering Research Council of Canada, and the Ontario Ministry for Research \& Innovation. Device fabrication benefited from the University of Waterloo's Quantum NanoFab, supported by the  Canada Foundation for Innovation, Industry Canada, and Mike \& Ophelia Lazaridis.

\section*{Appendix A: Additional characterization of device 1}

\begin{figure}[h]
\centering
\includegraphics[scale=1]{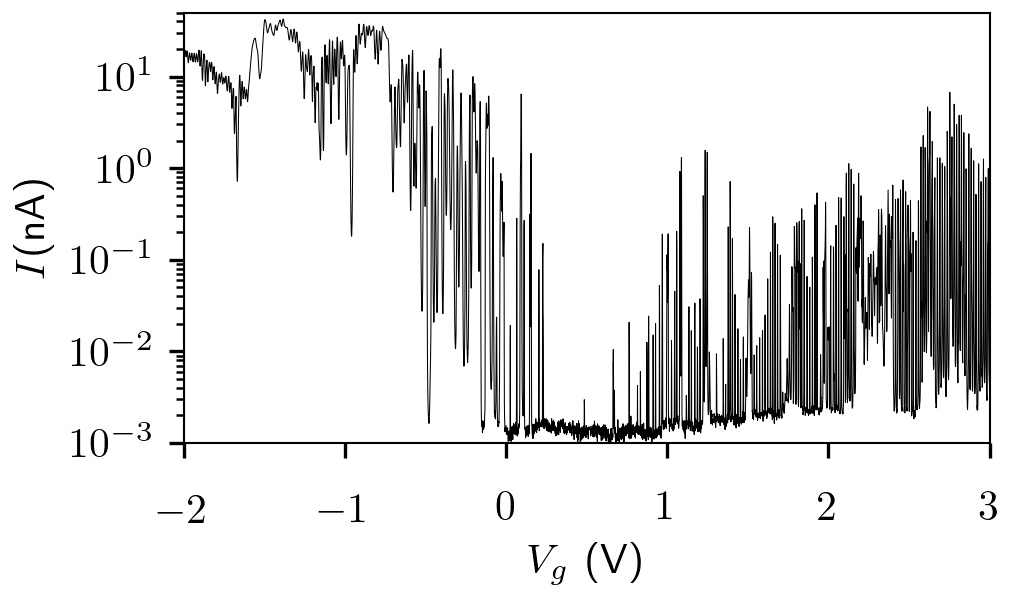}
\caption{Current as a function of applied gate voltage for the device discussed in the main text, measured at bias $V_{sd} = 1\:\text{mV}$. Large currents for $V_g < 0$ and Coulomb-blockaded conductance for $V_g > 0.5\:\text{V}$ are indicative of p-biased contacts to a single CNT.}
\label{fig:selfCond}
\end{figure}

Figure \ref{fig:selfCond} shows the current measured through device 1 as a function of gate voltage at applied bias of $V_{sd} = 1\:\text{mV}$. The bandgap region of suppressed current around $V_g = 0.4\:\text{V}$ and high conductance at $V_g < 0$ suggests hole biased metal-CNT interfaces and a bandgap of $E_{gap} \approx 80\:\text{meV}$. For $V_g > 0.5\:\text{V}$, the conductance is governed by Coulomb blockade.

\begin{figure}
\centering
\includegraphics[scale=1]{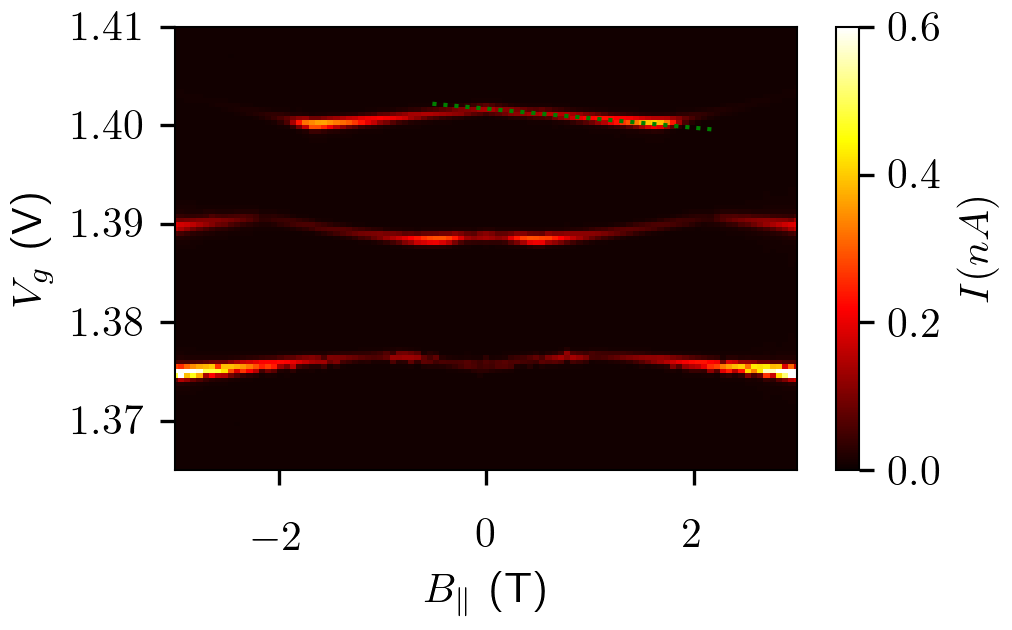}
\caption[Conductance peak position as a function of applied axial magnetic field in suspended CNT.]{Measured current through the CNT as a function of gate voltage and applied axial magnetic field, $B_\parallel$, at $V_{sd} = 0.5\:\text{mV}$. The conductance peak positions change with applied field due to orbital magnetic moments. Fitting the slope of the peak position change allows for a CNT diameter estimate. Using the slope indicated with the green dashed line, $d \approx 2\:\text{nm}$.}
\label{fig:orbMag}
\end{figure}

To estimate the CNT diameter, the axial field magneto-spectroscopy of three Coulomb peaks was measured, as shown in figure \ref{fig:orbMag}. The orbital magnetic moment is determined from the straight sections of these peak shifts, $\mu_{orb} = \alpha \frac{dV_{g,peak}}{dB_\parallel}$, where $\alpha \approx 0.45$ is the gate lever arm obtained from Coulomb diamonds. From the data in figure \ref{fig:orbMag}, an orbital magnetic moment of $\mu_{orb} = 0.4\:\text{meV/T}$ is calculated, which corresponds to a CNT diameter of approximately $2\:\text{nm}$ \cite{Ajiki1993}. 

\begin{figure}[h!]
\centering
\includegraphics[scale=1]{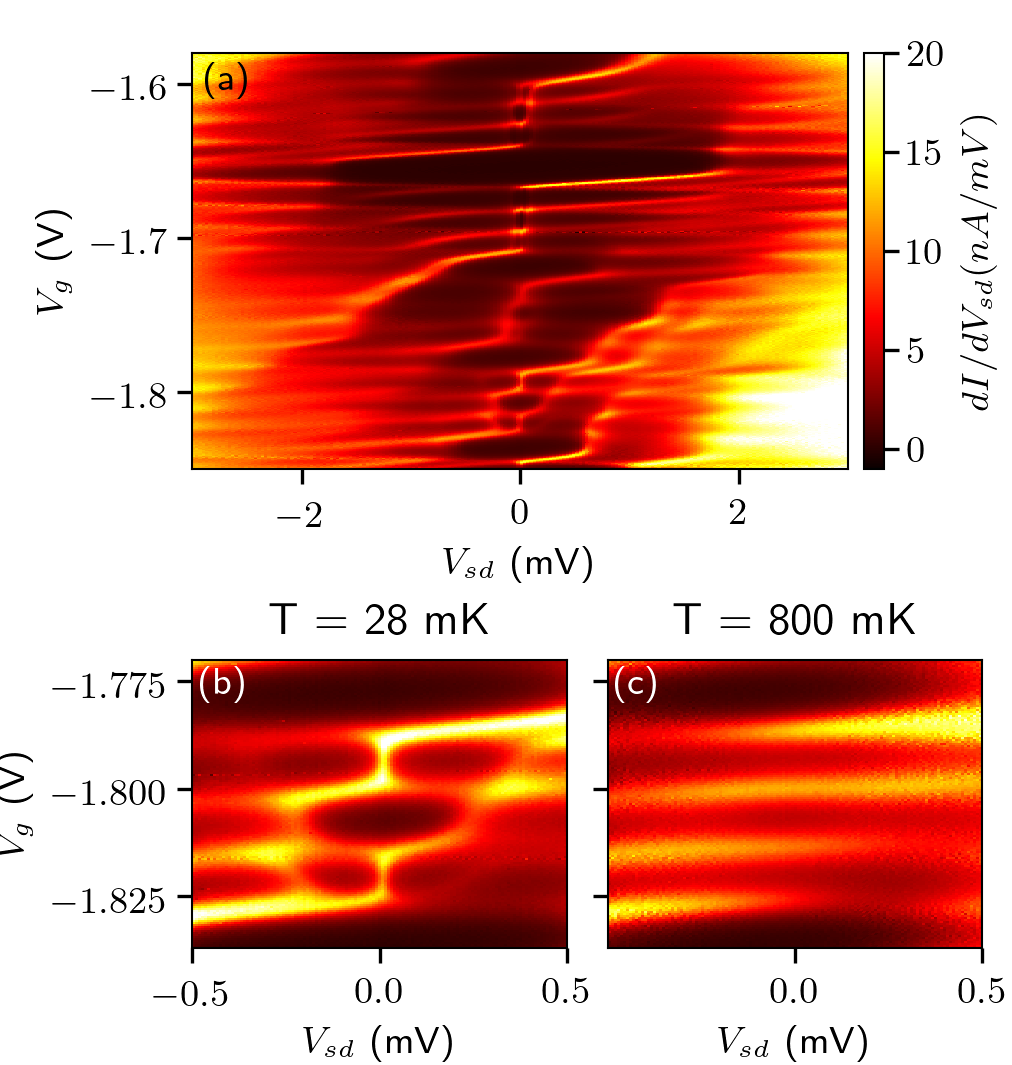}
\caption[Kondo effect in hole transport in the suspended CNT device]{Differential conductance measurements in hole-transport regime. \textbf{a} A wide area measurement showing four-fold shell filling over many charge occupations. \textbf{b} Focused measurement showing one set of four-shell filling, in which Kondo-enhanced transport at zero bias occurs for odd carrier occupation. \textbf{c} At $T = 800\:\text{mK}$, the transport is above the Kondo critical temperature, and the enhanced zero-bias transport is suppressed.}
\label{fig:selfKondo}
\end{figure}

Figure \ref{fig:selfKondo}a shows the differential conductance measured for a region of hole conduction ($V_g < 0$), which demonstrates four-fold shell filling and Kondo-enhanced transport \cite{Laird2014}. Figure \ref{fig:selfKondo}b shows a focus on one region of that data which demonstrates the Kondo-mediated transport, in which states with odd carrier occupation have enhanced zero bias conduction below the critical Kondo temperature. Figure \ref{fig:selfKondo}c shows the same measurement taken at temperature $T = 800\:\text{mK}$, where the Kondo enhanced transport has been suppressed. No self-driven oscillation features were observed in the hole transport at any gate voltage/bias.

\section*{Appendix B: Self-driven oscillation features at other gate voltages}

\begin{figure}[h]
\centering
\includegraphics[scale=1]{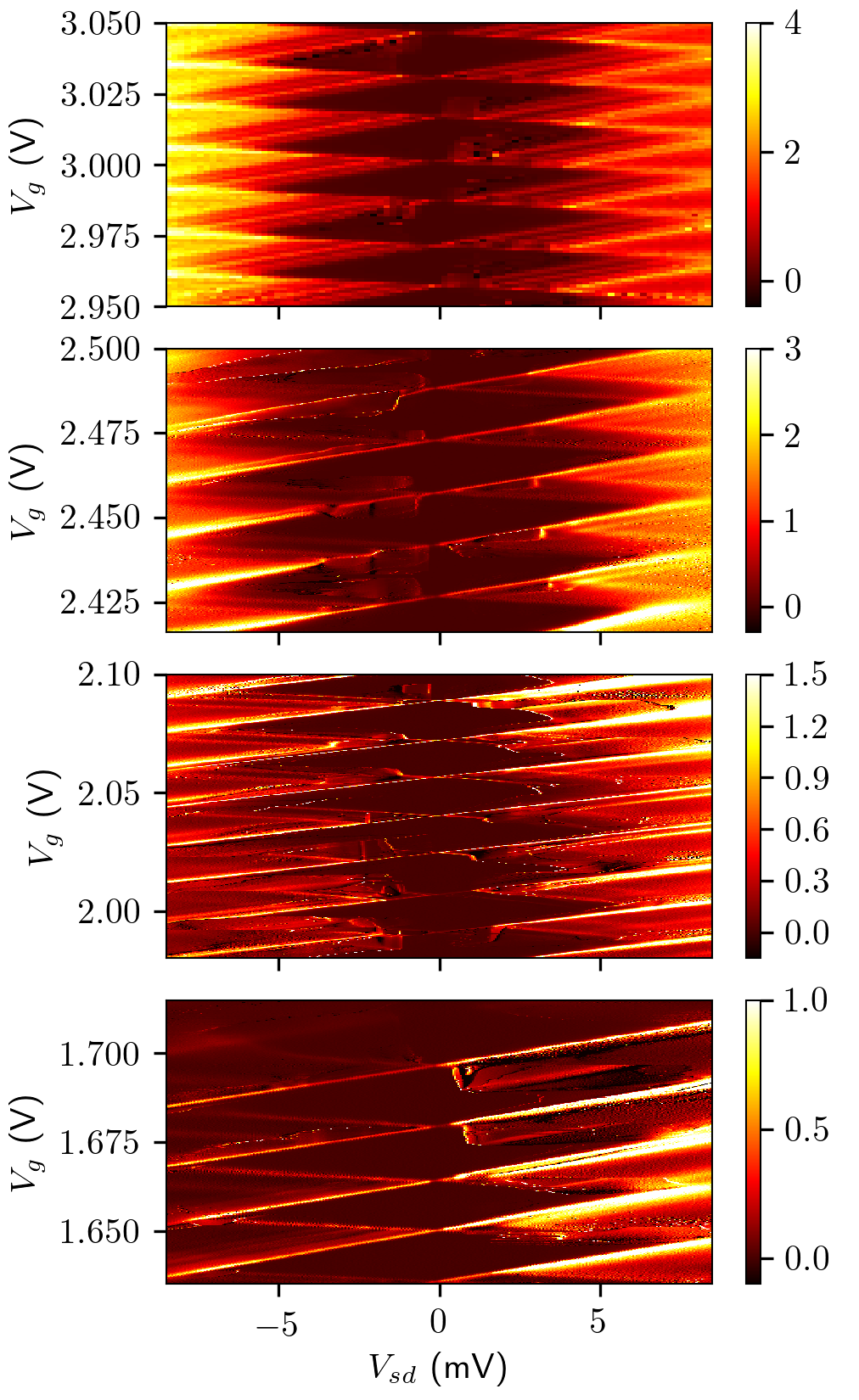}
\caption[Further example of in-diamond self-driven oscillation]{Differential conductance measurements as a function of gate voltage and applied bias, measured in device 1 at 30~mK. In each gate voltage range shown, finite conductance features similar to those discussed in the main text are observed. The location of these features within the Coulomb diamonds differ for each gate voltage range, indicating a change in the tunneling energy dependence. All color scales are given in $\mu\text{S}$.}
\label{fig:selfOtherDiamonds}
\end{figure}

The main text focused on the gate voltage range around $V_g = 1.8\:\text{V}$. Self-driven oscillation conductance features were also observed in the same device for other gate voltages in the electron conduction regime. Figure \ref{fig:selfOtherDiamonds} shows differential conductance measurements at several gate voltage ranges, each of which showed self-driving oscillation features within the nominally Coulomb blockaded diamond regions. Variation is observed in which Coulomb diamond edges are adjacent to the self-driven conductance features, indicating changes in the energy-dependent tunneling versus gate voltage.

\section*{Appendix C: Self-driven oscillations in device 2}

\begin{figure}[h]
\centering
\includegraphics[scale=1]{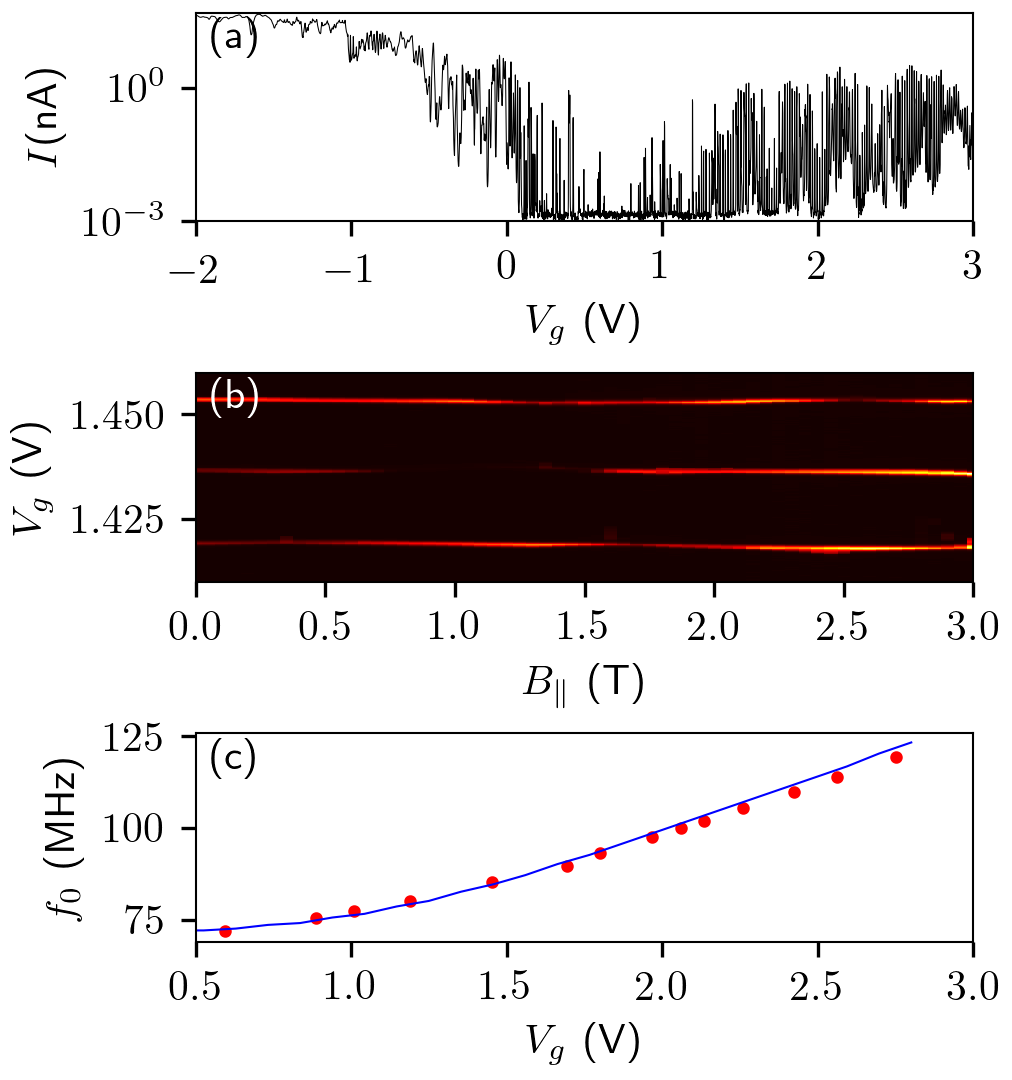}
\caption[Characterization of a second suspended CNT device which showed self-driven oscillations]{\textbf{a}. Current as a function of applied gate voltage for device 2. Similar to the device 1, this shows large currents for $V_g < 0$ and Coulomb-blockaded conductance for $V_g > 1\:\text{V}$. \textbf{b}. Conductance peak positions as a function of applied axial magnetic field, at $V_{sd} = 0.5\:\text{mV}$. The peak slopes give an orbital magnetic moment of $\approx 0.58\:\text{meV/T}$, corresponding to a CNT diameter of $2.9\:\text{nm}$. \textbf{c}. Resonance frequency as a function of gate voltage, and an Euler-Bernoulli beam model fit using CNT diameter of 2.9~nm, suspended length of $2.3\:\mu\text{m}$, and residual compression of $T_0 = 0.4\:\text{nN}$.}
\label{fig:devBbasics}
\end{figure}

In-diamond self-driven oscillations were also observed in a second suspended CNT device fabricated on the same chip, referred to as Device 2. Figure \ref{fig:devBbasics} shows electrical and mechanical characterization of this device. Similar to the device 1, device 2 shows p-biased metal-CNT interfaces and Coulomb-blockade for $V_g > 0.5\:\text{V}$. From the mechanical resonance frequency fitting shown in figure \ref{fig:devBbasics}c, the device is estimated to be a CNT of diameter 2.9~nm, suspended length of $2.3\:\mu\text{m}$, and a residual compression of $T_0 = 0.4\:\text{nN}$. Resonance peak linewidth gives $Q > 5 \cdot 10^3$. 

\begin{figure}
\centering
\includegraphics[scale=1]{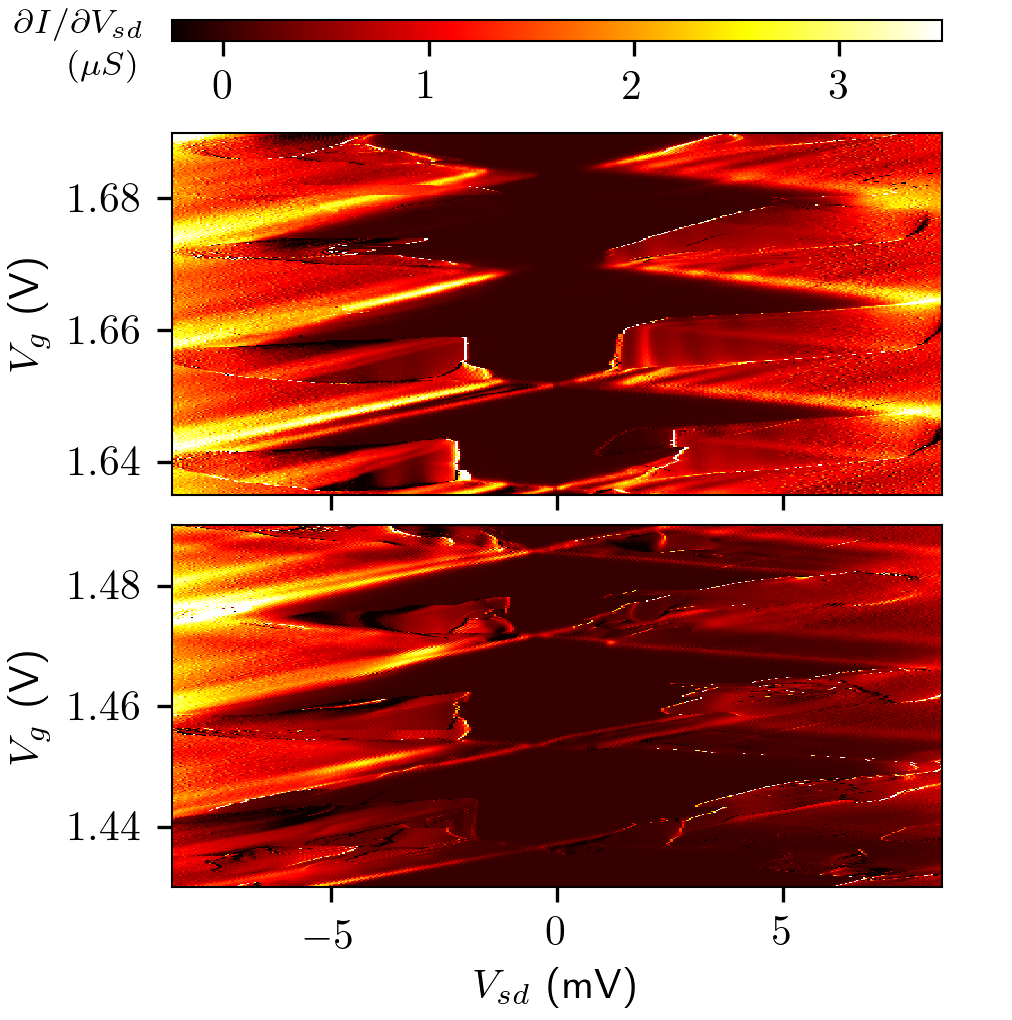}
\caption[In-diamond self-driven oscillations in device B]{Differential conductance measured in device 2. Self-driven oscillation features of finite conductance, qualitatively similar to those in device 1, are  observed within the nominally Coulomb blockaded diamond regions.}
\label{fig:selfDiamondB}
\end{figure}

Figure \ref{fig:selfDiamondB} shows differential conductance measurements at two different gate voltages in device 2, in which finite conductance self-driven oscillation features were observed in the Coulomb-blockade region.

\FloatBarrier

\end{document}